\documentclass [english,aps,prb,twocolumn,amsmath,amssymb,showpacs,showkeys]{revtex4}
\usepackage[T1]{fontenc}
\usepackage[latin1]{inputenc}
\usepackage{graphicx}
\usepackage{gensymb}
\makeatletter

\usepackage{epsfig}
\usepackage{prettyref}
\usepackage{babel}
\usepackage{subfigure}
\makeatother
\begin{document}
\title{Non-equilibrium effects in the magnetic behavior of Co$_{3}$O$_{4}$ nanoparticles}
\author{Vijay Bisht}
\email{vijayb@iitk.ac.in}
\author{K.P.Rajeev}
\email{kpraj@iitk.ac.in}
\affiliation {Department of Physics,
Indian Institute of Technology Kanpur 208016, India}

\begin{abstract}
We report  detailed studies on non-equilibrium magnetic behavior
of antiferromagnetic Co$_{3}$O$_{4}$ nanoparticles. Temperature
and field dependence of magnetization, wait time dependence of
magnetic relaxation (aging), memory effects and temperature
dependence of specific heat have been investigated  to understand
the magnetic behavior of these particles. We find that the system
shows some features characteristic of nanoparticle magnetism such
as bifurcation of field cooled (FC) and zero field cooled (ZFC)
susceptibilities and a slow relaxation of magnetization. However,
strangely, the temperature at which the ZFC magnetization peaks
coincides with the bifurcation temperature and does not shift on
application of magnetic fields up to 1~kOe, unlike most other
nanoparticle systems. Aging  effects in these particles are
negligible in both FC and ZFC protocol and memory effects are
present only in FC protocol. We estimate the Néel temperature by
using Fisher's relation as well as  directly by measurement of
specific heat, thus testing the validity of Fisher's relation for
nanoparticles. We show that Co$_{3}$O$_{4}$ nanoparticles
constitute a unique aniferromagnetic system which develops a
magnetic moment in the paramagnetic state because of
antiferromagnetic correlations and enters into a blocked state
above the Néel temperature.

\end{abstract}
\pacs{75.50.Tt,75.20.-g,75.75.Jn,75.50.Lk}
\keywords{Co$_{3}$O$_{4}$ nanoparticles, magnetic properties,
superparamagnetism, memory, aging, antiferromagnetic
correlations,Fisher's relation, FC-ZFC.}

\maketitle

\section{INTRODUCTION}

In the past two decades, magnetic nanoparticles have attracted
much attention due to (a) their potential uses in various areas
such as data storage and biomedicine and (b) the challenge to
understand the physics underlying the various exotic phenomena
exhibited by them.\cite{Dormann,Steen,Pankhurst} Most of such work
has focussed on ferromagnetic and ferrimagnetic nanoparticles
because of their high magnetic moments that make them industrially
valuable. However, there have been a few studies on
antiferromagnetic nanoparticles and interestingly, their magnetic
behavior is found to be more complex and intriguing.

 Below a critical size, particles of a ferromagnetic or ferrimagnetic
material consist of a single domain and each particle carries a
magnetic moment, which can reverse its direction due to thermal
agitation. In these particles, the magnetic dynamics is thermally
activated leading to paramagnetic behavior above a particular
temperature, called blocking temperature. This phenomenon is known
as superparamagnetism as instead of atomic spins, particle moments
('superspins') are involved. Below this temperature, the moments
appear frozen in a particular direction, on the time scale of the
measurement. Antiferromagnetic nanoparticles can also develop a
net moment due to uncompensated spins and can exhibit
superparamagnetism as proposed by Néel.\cite{Neel,Brown} Thus
magnetic nanoparticles are expected to show superparamagnetism,
though, interparticle interactions can complicate matters leading
to spin glass like
behavior.\cite{Bisht,Batlle,Sasaki,Sun,Malay,Sahoo} In addition to
interparticle interactions, spin glass behavior can arise within a
particle due to freezing of spins at the
surface.\cite{Tiwari,Martinez, Nadeem}

Magnetic nanoparticles show some characteristic features which are
common to both spin glasses and superparamagnets. These include
irreversibility in the field cooled (FC) and zero field cooled
(ZFC) magnetizations, a peak in ZFC magnetization, slow magnetic
relaxation and hysteresis at low temperature. However, some
important features that distinguish these two are: (1) FC
magnetization goes on increasing as the temperature is decreased
in superparamagnets while it tends to saturate below the peak
temperature in particles showing spin glass behavior.\cite{Sasaki}
(2) In systems showing spin glass like behavior, wait time
dependence of relaxation (aging) and memory effects are present in
both FC and ZFC magnetizations. In contrast, in superparamagnetic
particles, these effects are present in FC magnetization
only.\cite{Bisht,Batlle,Suzuki,Sasaki,Sun,Malay,Chakraverty,
Tsoi,RZheng,Martinez,Sahoo} (3) The field dependence of
temperature at which the ZFC magnetization peaks
($T_{\textnormal{P}}$) is known to behave differently in the two
cases.\cite{Tiwari,Almeida,Bitoh,RKZheng}

There have been studies on various antiferromagnetic nanoparticles
in both bare and coated forms in the past several
years.\cite{Steen} It has been noticed that there is a lot of
variation in the magnetic behavior of the nanoparticles of
different materials. For example, NiO nanoparticles have been
reported to show spin glass like behavior; \cite{Bisht,Tiwari} CuO
nanoparticles show an anomalous magnetic behavior that cannot be
described as superparamagnetic or spin glass like; \cite{Vijay}
Ferritin shows superparamagnetic behavior \cite{Kilcoyne,Sasaki}
etc. Co$_{3}$O$_{4}$ is an antiferromagnetic  material and in the
bulk form, its  Néel temperature has been reported to lie between
30~K and 40~K.\cite{Resnick}There have been some reports on
hysteresis, time dependence of magnetization, exchange bias and
finite size effects in bare, coated and dispersed Co$_{3}$O$_{4}$
nanoparticles and various claims have been made in support of spin
glass like and superparamagnetic behavior in these particles.
\cite{Yuko,Takada,Yuko1,Resnick,Makhlouf,Lin,Michi,
Mousavand,Shandong,Dutta}  It will be, therefore, worthwhile to
investigate their magnetic behavior carefully. In the present
work, we  present a detailed study on non-equilibrium features
such as temperature, time and field dependence of magnetization,
aging and memory effects. We also report specific heat
measurements to find the Néel temperature ($T_{\textnormal{N}}$)
of Co$_{3}$O$_{4}$ nanoparticles. Usually $T_{\textnormal{N}}$ is
estimated in nanoparticles using the Fisher's equation that
relates specific heat (C) and magnetic susceptibility ($\chi $) of
antiferromagnetic materials.\cite{Dutta,Fisher} We determine
$T_{\textnormal{N}}$ using both methods and are able to test the
validity of
 Fisher's relation in these nanoparticles.

\section{EXPERIMENTAL DETAILS}

Co$_{3}$O$_{4}$ nanoparticles are prepared by a sol gel method.
Aqueous solution of sodium hydroxide is mixed with that of cobalt
nitrate till the pH of the solution becomes 12. At this stage
cobalt hydroxide  separates from the solution forming a gel that
is centrifuged to obtain a precipitate which is washed with water
and ethanol several times and  dried to obtain a precursor. This
precursor is heated at 250 $^0$C for 3 hours to obtain
Co$_{3}$O$_{4}$ nanoparticles. The sample is characterized by
X-ray diffraction (XRD) using a Seifert diffractometer with
Cu~K$\alpha$ radiation and Transmission electron microscopy (TEM)
using  FEI Technai 20 U Twin Transmission Electron Microscope. All
the magnetic measurements are done with a SQUID magnetometer
(Quantum Design) and specific heat measurements are done with a
PPMS (Quantum Design).

\section{RESULTS AND DISCUSSION}

\subsection{Particle size}
\begin{figure}[t]
\begin{centering}
\includegraphics[width=1\columnwidth]{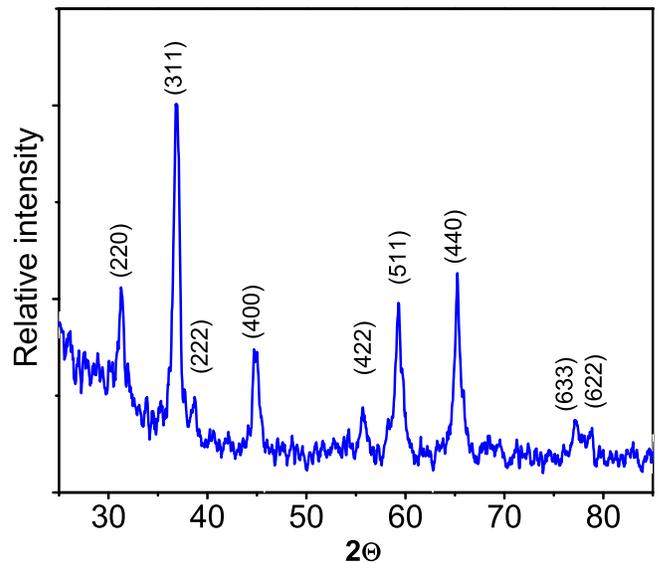}
\par\end{centering}
\caption{(Color online) XRD pattern of sample heated at
$250\degree C$ for 3 hours. All the peaks correspond to those of
Co$_3$O$_4$. } \label{fig:XRD}
\end{figure}

\begin{figure}[b]
\begin{centering}
\includegraphics[width=1\columnwidth]{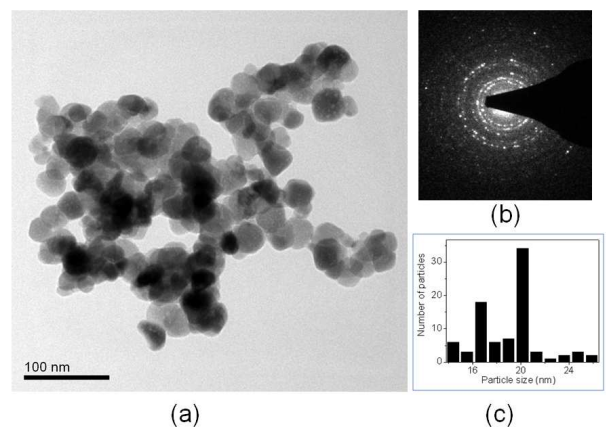}
\par\end{centering}
\caption{(a) TEM image of the sample. (b) Selected area
diffraction (SAD) pattern. (c) Histogram of the particle size
distribution. Total Number of particles considered is 84.}
\label{fig:TEM}
\end{figure}

The XRD pattern of the sample (Figure \ref{fig:XRD}) corresponds
to that of Co$_3$O$_4$. The average particle size as estimated
from the broadening of XRD peaks using the Scherrer formula turns
out to be $12.5~nm$.  TEM image is shown in Figure \ref{fig:TEM}
(a). Parts (b) and (c) of this  figure show the corresponding
selected area diffraction (SAD) pattern and  the particle size
distribution of the sample. It can be seen that the particles are
more or less spherical in shape and  the average particle size
estimated comes out to be about 18 nm with a standard deviation of
1.5 nm. The SAD pattern consists of concentric diffraction rings
with different radii. The diameter of a diffraction ring in SAD
pattern is proportional to \(\sqrt{h^{2} + k^{2} + l^{2}}\), where
\(({h}{k}{l})\) are the Miller indices of the planes corresponding
to the ring. Counting the rings from the center 1st, 2nd, 3rd,...
rings correspond to (220), (311), (222),.. planes, respectively,
in agreement with the XRD pattern.

\subsection{Temperature and field dependence of magnetization}

\begin{figure}[t]
 \begin{centering}
\includegraphics[width=1\columnwidth]{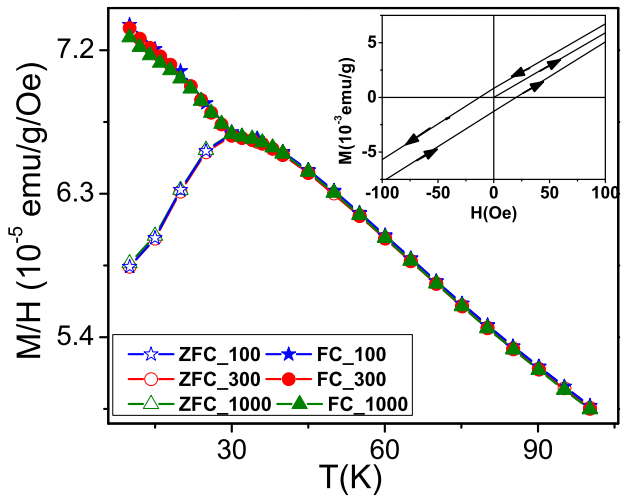}
\par\end{centering} \caption{(Color online) Temperature dependence
of FC and ZFC susceptibilities  at 100~Oe, 300~Oe and 1000~Oe.
Inset shows the hysteresis at 5~K.}
 \label{fig:FCZFC}
\end{figure}

The temperature dependence of magnetization was done under field
cooled (FC) and zero field cooled (ZFC) protocols at fields
100~Oe, 300~Oe and 1000~Oe. See Figure \ref{fig:FCZFC}. We note
that the FC magnetization in this case is increasing with decrease
in temperature down to the lowest temperature of measurement
without any sign of saturation, a feature characteristic of
superparamagnets.\cite{Sasaki} However, the temperature at which
the ZFC magnetization peaks (31~K) i.e. $T_{\textnormal{P}}$ is
also the temperature of bifurcation ($T_{\textnormal{bf}}$) of FC
and ZFC magnetizations.\cite{footnote} Strangely, in this case,
($T_{\textnormal{P}}$) is independent of the magnetic field
applied and the dc susceptibility data taken at different fields
superpose. This is in contrast to other nanoparticles where the
peak temperature is found to shift to lower temperatures on
increasing the field even at fields as low as a few hundred
Oeresteds.\cite{Bitoh,RKZheng,Almeida,Tiwari,Kachkachi,Sappey}
Another distinct feature in Figure \ref{fig:FCZFC} is a sudden
change in the slope of FC magnetization curve at the peak
temperature. This will be discussed later in subsection D.

 We have done hysteresis measurements at 5~K and a magnified view
 is shown in the inset of Figure \ref{fig:FCZFC}. The virgin curve
 in this case is a straight line in contrast to spin glasses where
 it is usually S-shaped.\cite{Mydosh}

\begin{figure}[b]
\begin{centering}
\includegraphics[width=1\columnwidth]{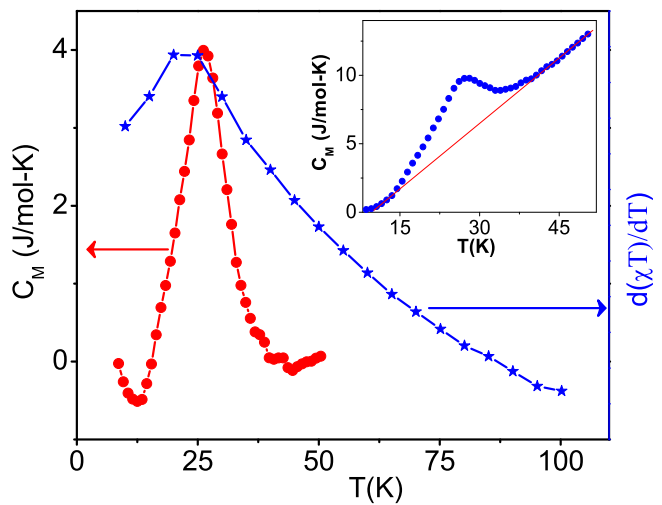}
\par\end{centering}
\caption{(Color online) Magnetic contribution to specific heat
estimated by subtracting the linear component at lower
temperatures and plot of  $d(\chi T)/dT$ as a function of
temperature. Inset shows the plot of specific heat as a function
of temperature. The line (red) shows the linear part of specific
heat which is subtracted from the total specific heat to get the
magnetic contribution.} \label{fig:specheat}
\end{figure}

\subsection{Néel temperature}

When the particle size of an antiferromagnetic material is
decreased to nanoscale, its Néel temperature
($T_{\textnormal{N}}$) is known to decrease significantly due to
finite size effects.\cite{Zheng,Lin} We carried out specific heat
measurements on pelletized  nanoparticles to make an estimate of
$T_{\textnormal{N}}$ and these data are presented in the inset of
Figure~\ref{fig:specheat}. It can be observed that the specific
heat decreases linearly with decrease in temperature down to 35~K
below which it increases with a peak at 28~K. We calculated the
magnetic specific heat  by subtracting a linear contribution
 from  the total specific heat and this data is shown in the
  main panel of Figure~\ref{fig:specheat}. It can
be seen that a magnetic contribution is noticeable between 15~K
and 40~K with a peak at 26~K, which should correspond to the Néel
temperature.

 In some works on antiferromagnetic nanoparticles,
$T_{\textnormal{N}}$ has been identified as the temperature of the
peak of $ d (\chi T)/dT$ vs $T$ curve, where $\chi$ is the dc
susceptibility(ZFC).\cite{Dutta,Punnoose} This is in accordance
with Fisher's equation relating specific heat (C) and magnetic
susceptibility ($\chi $) of antiferromagnetic
systems:\cite{Fisher}

\begin{equation} \label{Fisher relation} C \propto  d
(\chi T)/dT \end{equation}
 This relation has been verified
experimentally for some bulk materials.\cite{Bragg} It will be
interesting to check the validity of this relation for the present
nanoparticle sample. In Figure \ref{fig:specheat}, we have also
shown the plot of $d(\chi T)/dT$ as a function of $T$, which has a
peak somewhere in between 20~K to 25~K, giving an estimate of the
Néel temperature. Thus the value of $T_{\textnormal{N}}$ extracted
from the specific heat is somewhat greater than the value obtained
from the Fisher relation.

 \subsection{Aging and memory effects}

Nanoparticles that show spin glass like behavior are expected to
show aging and memory effects in both FC and ZFC protocols while
those showing superparamagnetic behavior  show these effects only
in FC protocol. Further the effects in FC protocol are weaker in
superparamagnetic particles than in the systems showing spin glass
like behavior.\cite{Bisht,Batlle,Sasaki} These experiments can
thus give valuable information about the nature of magnetic
behavior of a system.

\begin{figure}[t]
\begin{centering}
\includegraphics[width=1\columnwidth]{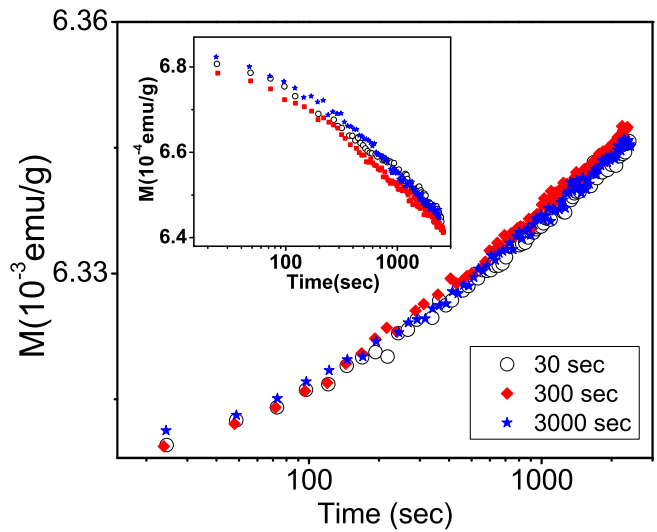}
\par\end{centering}
\caption{(Color online) Aging experiments in ZFC protocol at 20~K
with waiting times 30~s, 300~sec and 3000~sec. Inset shows the
corresponding experiments in FC protocol. } \label{fig:aging}
\end{figure}

For doing aging experiments in FC protocol, the sample is cooled
to a particular temperature in a field of 100~Oe, the field is
switched off after waiting for a specified period and
magnetization is recorded as a function of time.  In the
corresponding ZFC case, the sample is cooled to the temperature of
interest in zero field and the field is switched on after a
certain wait time and subsequently magnetization is recorded as a
function of time. We show the results of aging experiments at 20~K
in Figure~\ref{fig:aging}. It can be seen that aging effects are
negligible in both FC and ZFC measurements.

\begin{figure}[b]
\begin{centering}
\includegraphics[width=1\columnwidth]{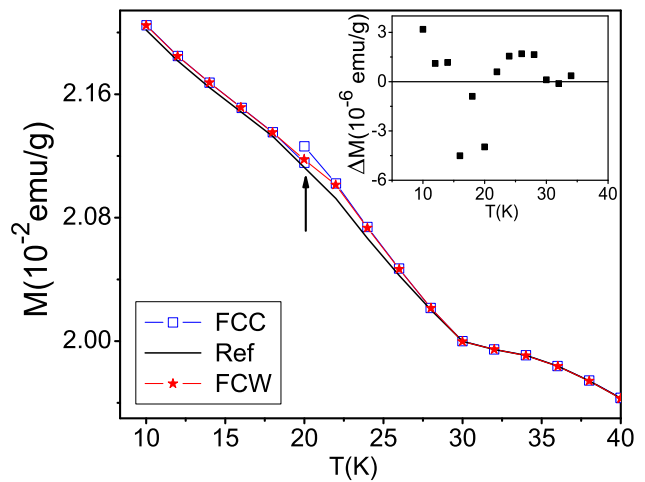}
\par\end{centering}
\caption{(Color online) Memory experiments in FC protocol with a
stop of one hour duration at 20~K at a field of 300~Oe. The field
is switched off during the stop as indicated by the arrow. Inset
shows the corresponding data for ZFC memory experiment. The
difference in magnetization with a stop of one hour at 20K in the
cooling process and the reference data is plotted as a function of
temperature. } \label{fig:FCmemory}
\end{figure}

 For carrying out FC memory experiments, the system is cooled in
 the presence of a magnetic field to 20~K
and a stop of one hour is taken at this temperature. Magnetic
field is switched off for the duration of the stop and is turned
on before  cooling it further to 10~K. The magnetization is
measured while cooling and then during subsequent heating. These
data have been taken at  300~Oe field. See
Figure~\ref{fig:FCmemory}. It can be observed that there is some
indication of  memory as the heating curve meets the cooling curve
just above the temperature at which the stop was taken.  We have
also done memory experiments in ZFC protocol with a stop of one
hour at 20~K. For doing these measurements,
 we first cool the sample
  continuously down to 10~K in zero field and applying a field of
  300~Oe to record the magnetization data as the temperature is
   increased in steps up to 100~K. We call this data as the
ZFC reference. Now the sample is cooled in  zero field down to
10~K with a stop of one hour at 20~K and during subsequent heating
the magnetization is recorded with an applied field of 300~Oe as
the temperature is increased to 100~K. In the inset of
Figure~\ref{fig:FCmemory}, we show $\Delta M $, the difference in
magnetization between the ZFC data with the stop and the ZFC
reference, as a function of temperature. We observe that there is
no indication of memory at the temperature at which stop was taken
in the cooling process and $\Delta M $ is less than $0.05\%$. This
is in contrast to canonical spin glasses or nanoparticle systems
that show spin glass like behavior where a clear dip in $\Delta M
$ vs $T$ curve is observed at the temperature where the halt was
made during the cooling process and in the vicinity of the dip
$\Delta M $ is of order $1\%$ or more.\cite{Bisht,Matheiu} From
these experiments, it is clear that below the peak temperature,
the irreversibility in FC and ZFC magnetization is  due to
superparamagnetic blocking.


\subsection{DISCUSSION}

We have seen that Co$_{3}$O$_{4}$ nanoparticles show some features
characteristic of nanoparticle magnetism. Some of these are due to
finite size effects viz. a net magnetic moment due to
uncompensated surface spins and a decrease in Néel temperature.
Some are manifestations of non equilibrium in magnetic
nanoparticles viz. irreversibility in FC and ZFC magnetization and
a slow magnetic relaxation at low temperature. However there are
several unusual features observed in this system, that deserve a
closer look.

\subsubsection{Behavior above $T_{\textnormal{P}}$}

\begin{figure}[!t]
\begin{centering}
\includegraphics[width=0.85\columnwidth]{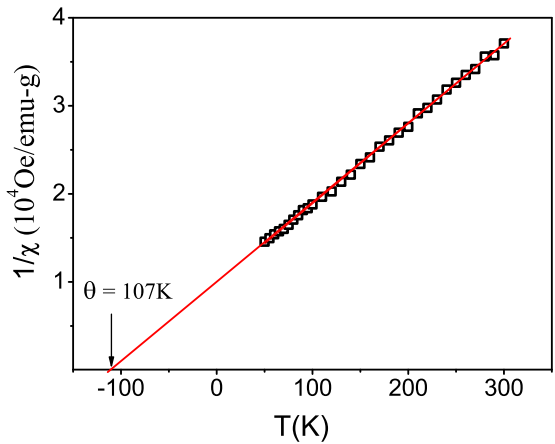}
\par\end{centering}
\caption{(Color online)Plot of $1/\chi$ vs $T$ for the temperature
range 50~K to 300~K for ZFC magnetization taken at an applied
field  200 Oe. Solid line (red) shows the linear fit to the data.
} \label{fig:AFMfit}
\end{figure}

In ferromagnetic and ferrimagnetic nanoparticles, the ZFC
magnetization generally shows a peak at a particular temperature,
above which the behavior is superparamagnetic. This behavior is
characterized by  the superposition of magnetization curves taken
at various temperatures above $T_{\textnormal{P}}$  when plotted
against $H/T$.\cite{Bean} However, Makhlouf et al. have shown that
in Co$_{3}$O$_{4}$ nanoparticles, magnetization vs $
H/(T$+$\theta$) curves taken  at temperatures above 50~K superpose
with ($\theta = 85~K$), a feature characteristic of
antiferromagnetic materials.\cite{Makhlouf}  Our data (50~K-300~K)
also fits well to Curie-Weiss law, $\chi \propto 1/(T+\theta)$,
with $\theta = 107~K$ and coefficient of determination, $R^2 =
0.9992$. See Figure ~\ref{fig:AFMfit}. Thus, the system is not in
a superparamagnetic state above $T_{\textnormal{P}}$. It may be
noted that $T_{\textnormal{N}}$ as found by specific heat
measurements is 5~K less than $T_{\textnormal{P}}$ and the Fisher
relation gives an even lower estimate of $T_{\textnormal{N}}$.


\subsubsection{Behavior below  $T_{\textnormal{P}}$}

 There have been some reports of spin glass like features in Co$_{3}$O$_{4}$
nanoparticles coated with organic surfactants and those dispersed
in  amorphous matrices at low temperature.
\cite{Resnick,Michi,Mousavand,Shandong} However, in the present
work and in other studies on bare nanoparticles, no such features
have been found. \cite{Makhlouf,Dutta} Absence of aging and memory
effects in ZFC protocol confirms that the behavior of
Co$_{3}$O$_{4}$ nanoparticles is not spin glass
like.\cite{Bisht,Sasaki,RZheng,Tsoi} Thus below
$T_{\textnormal{P}}$, observation of a bifurcation in FC and ZFC
magnetization and a slow magnetic relaxation seems to correspond
to a blocked state as observed in superparamagnetic particles.
Presence of memory effects in FC protocol also support this
inference.

At $T_{\textnormal{P}}$, there also occurs a bifurcation between
FC and ZFC magnetization and this change looks abrupt as at this
point a slope change in the FC magnetization can be seen. Thus
even before the actual transition to an antiferromagnetic state,
the particles get blocked as indicated by the FC and ZFC
bifurcation.  We propose that this is due to the short range
antiferromagnetic correlations which are known to persist above
the Néel temperature.\cite{buschow,Vijay}
 These short ranged correlations can give rise to a net magnetic
 moment when the correlation length becomes comparable to the particle
 size and thus the particles can get blocked above
 $T_{\textnormal{N}}$.


   \vspace{1 cm}

\section{Conclusion}

  We have done a detailed study of non-equilibrium  magnetic behavior of
Co$_{3}$O$_{4}$ nanoparticles. We find that their behavior is
unique among antiferromagnetic nanoparticles. There is a peak in
ZFC magnetization  and at this temperature ($T_{\textnormal{P}}$),
a sudden bifurcation between FC and ZFC magnetization occurs,
strangely, above the Néel temperature. There is a sudden slope
change in FC magnetization at the bifurcation temperature and the
magnetization keeps on increasing on decreasing the temperature, a
feature characteristic of superparamagnetic nanoparticles.
However, the behavior of susceptibility above the peak temperature
is paramagnetic rather than superparamagnetic. Aging and memory
effects are not observed in ZFC magnetization measurements which
show the absence of spin glass like behavior in this system.
However observation of memory in FC protocol supports
superparamagnetic blocking below $T_{\textnormal{P}}$. Thus, above
$T_{\textnormal{P}}$, Co$_{3}$O$_{4}$ nanoparticles are
paramagnetic with antiferromagnetic correlations and below
$T_{\textnormal{P}}$, they get blocked  in the paramagnetic state
itself even before the transition to an antiferromagnetic state at
$T_{\textnormal{N}}$.

 \begin{acknowledgments}
VB thanks the University Grants Commission of India for financial
support. Authors thank Prof. C .V .Tomy, IIT Bombay for specific
heat measurements.
\end{acknowledgments}


\begin{thebibliography}{40}
\bibitem{footnote} Some authors have observed a $T_{bf}$
which lies a few Kelvins above $T_P$ and some others have found it
even below $T_P$. \cite{Dutta,Makhlouf}
\end{thebibliography}
\end{document}